\documentclass[preprint,final,5p,times,twocolumn]{elsarticle}
\usepackage{rotating,color,subfigure,amssymb}
\usepackage{alphalph}\renewcommand\theaffn{\alphalph{\arabic{affn}}}
\usepackage{amsmath}
\usepackage[T1]{fontenc}
\usepackage{amssymb}
\journal{Physics Letters B}
\begin{document}
                                       
\begin{frontmatter}

\title{The $d^*(2380)$ in neutron stars - a new degree of freedom?}
\date{\today}

\author[INFN]{I.~Vida\~na}
\author[UoE]{M.~Bashkanov}\ead{mikhail.bashkanov@ed.ac.uk}
\author[UoE]{D.P.~Watts}   
\author[UoY]{A.~Pastore} 
\address[INFN]{INFN Sezione di Catania. Dipartimento di F\'{i}sica, Universit\`a di Catania, Via Santa Sofia 64, 95123 Catania, Italy}
\address[UoE]{School of Physics and Astronomy, University of Edinburgh, James Clerk Maxwell Building, Peter Guthrie Tait Road, Edinburgh EH9 3FD,UK}
\address[UoY]{Department of Physics, University of York, Heslington, York, Y010 5DD, UK}

\cortext[coau]{Corresponding author }

\begin{abstract}
Elucidating the appropriate microscopic degrees of freedom within neutron stars remains an open question which impacts nuclear physics, particle physics and astrophysics. The recent discovery of the first non-trivial dibaryon, the $d^*(2380)$, provides a new candidate for an exotic degree of freedom in the nuclear equation of state at high matter densities. In this paper a first calculation of the role of the $d^*(2380)$ in neutron stars is performed, based on 
a relativistic mean field description of the nucleonic degrees of freedom supplemented by a free boson gas of $d^*(2380)$. The calculations indicate that the $d^*(2380)$ would appear at densities around three times normal nuclear matter saturation density and comprise around 20\% of the matter in the center of heavy stars with higher fractions possible in the higher densities of merger processes. The $d^*(2380)$ would also reduce the maximum star mass by around 15\% and have significant influence on the fractional proton/neutron composition. New possibilities for neutron star cooling mechanisms arising from the $d^*(2380)$ are also predicted. 
\end{abstract}

\end{frontmatter}

\section{Introduction}

Neutron stars are valuable laboratories to study the fundamental properties of dense nuclear matter at low temperatures. Despite the major advances in neutron star astronomy in recent years, significant gaps remain in our knowledge of their internal structure. The appropriate microscopic degrees of freedom are only established for densities comparable to atomic nuclei, which are found in the neutron star crust or at shallow depths in typical stars. The properties of this nucleonic matter can be constrained by precision measurements of nuclei and nuclear reactions. This has led to a flurry of recent theoretical~\cite{steiner,Horowitz,Zu,Steiner2,rutel} and experimental~\cite{Tsang2,cohpi,prex} nuclear physics programmes to improve constraints on the equation of state for pure nucleonic matter. 

However, for higher densities existing in the cores of heavy neutron stars or in neutron star mergers~\cite{NSML} ``exotic'' phases beyond simple nucleonic matter are postulated.  A number of possibilities have been investigated theoretically, such as pion condensates \cite{pion}, kaon condensates \cite{kaon}, hyperonic matter \cite{EPJA16}, $\Delta$ isobars~\cite{Drago1,Drago2}, quark matter \cite{quark}  and dibaryonic matter~\cite{NM3}. Some of the proposed exotic phases were resolved and rejected.  The astronomical observation of heavy neutron stars with $\sim 2 M_{\odot}$~\cite{Star1,Star2,Abbott}, places stringent constraints on the contribution of exotic components. Any extra degree of freedom, in addition to protons, neutrons and leptons tends to make the nuclear Equation of State (EoS) softer, consequently reducing the maximum stable mass for a neutron star. Apparent inconsistencies with the observed neutron star properties when including some of the exotic phases has led to ``puzzles'' {\it e.g.,}  the ``hyperon puzzle'' \cite{EPJA16}  or the ``$\Delta$ puzzle'' \cite{Drago2}. However, such exotic phases are generally in competition and the inclusion of one process may strongly influence the role of the others. To reach firm conclusions on the microscopic composition of dense nuclear matter a comprehensive investigation of all potential processes appears crucial. Achieving this goal would be  very timely. Neutron star mergers, such as recently observed by LIGO with both electromagnetic and gravitational wave signals, have properties which depend strongly on the EoS \cite{Abbott} and are now proposed as the main site of galactic heavy element production. A better understanding of matter at high densities is also needed for studies of black hole formation \cite{bh} and neutron star black-hole mergers \cite{ruffert}.

Dibaryonic degrees of freedom have the potential for major impact on the properties of the dense nuclear matter.  The existence of dibaryons has long been predicted by phenomenological models based on the theory of the strong force, quantum chromodynamics (QCD). Dibaryons are predicted to be colourless QCD objects comprising 6 valence quarks and having integer spin.  In the quantum mechanical environment within a neutron star a dibaryon would essentially become a stable particle and its bosonic nature would lead to a very different physical behaviour than for the (fermionic) nucleons. For example, the dibaryon can form into a Bose-Einstein condensate~\cite{TAM}.  This, as well as the small predicted size of the dibaryons (6 quarks occupying similar size to a single nucleon~\cite{DBT9}) indicate they could be a new way to distribute energy in matter under compression. 

Due to the potential significance of dibaryons in understanding the properties and structure of neutron stars it is very important to explore the influence of the first non-trivial dibaryon, the $d^*(2380)$ recently discovered by the Wasa-at-Cosy collaboration. In this work we perform the first evaluation of the effect of this specific $d^*(2380)$ dibaryon on the nuclear EoS.  The paper will firstly summarise the evidence for the $d^{*}(2380)$ and then present the first theoretical investigation of its influence on neutron star properties.


\section{The $d^*(2380)$ dibaryon}
Theoretical predictions of dibaryons, such as the $d^{*}$, have a long history. Calculations started with the pioneering work of
Dyson and Xuong ~\cite{DYS} in 1964 and have developed in many studies since then~\cite{BBC, AG1,AG2,AG3,FW,QM2,DBT1,DBT2,DBT3,DBT4,DBT5,DBT6,DBT7,DBT8,DBT9,DBT10}. 
 Despite the use of different ansatz in the models (constituent quarks, $\Delta\Delta$ molecule,  compact hexaquark with a $\sigma$-meson cloud) the majority of the models indicate the $d^*$ is a compact object, $I(J^P) = 0(3^+)$ with a size comparable to that of a single nucleon or somewhat larger in the case of molecular picture. Calculation of the $d^*$ dibaryon properties directly from the QCD lagrangian is ongoing but challenging. First Lattice QCD calculations indicate very strong $\Delta\Delta$ attraction where their spins are aligned, as expected from a $d^*$ ~\cite{LQCD}.  
 
Experimentally, a solid candidate for the $d^{*}$ has only emerged in recent years. Measurement of the basic double-pionic fusion reactions $pn \to d \pi^0\pi^0$ and $pn \to d \pi^+\pi^-$ revealed a narrow  resonance-like structure in the integrated cross section~\cite{mb,MB,MBC,Kuk} at a mass $M \approx$ 2380~MeV with unusually narrow width of $\Gamma \approx$ 70 MeV. The data are consistent with a $I(J^P) = 0(3^+)$ assignment for this resonant structure. Additional evidence has been obtained in the $pn \to pp\pi^0\pi^-$\cite{TS1}, $pn \to pn\pi^0\pi^0$~\cite{TS2}  and $\vec n p \to d \pi^0\pi^0$~\cite{MBA}  reactions. Partial wave analysis including new polarised $ \vec n p$ elastic scattering data~\cite{MBE1,MBE2,WRK} confirms the existence of a resonance pole at $(2380\pm 10)-i(40\pm 5))$ MeV. For a recent review of dibaryon searches see Ref.~\cite{Cl1}.

The $d^*(2380)$ is the only exotic particle which can be produced copiously at modern experimental facilities. Its basic properties and all major decay branches have therefore been  determined in a short period of time~\cite{BCS}. All the data ~\cite{MBE1,MBE2, mb, MB, MBC, TS1, TS2, TS3} collected so far suggest that in 88 percent of cases the $d^*(2380)$ decays into $\Delta\Delta$ and in 12\% to $pn$, with high angular momentum $L=2$ or $4$ between nucleons~\cite{BCS,PBC,BCS1}. Further $d^*(2380)$ studies indicate it also has an electromagnetic coupling~\cite{Toh,Bas,TOK1,TOK2,EDPOL}. 

From measurements of  $pd \to ^3$He~$\pi\pi$ ~\cite{MBH,EP}, $dd \to ^4$He~$\pi\pi$ ~\cite{AP} and heavy ion collisions~\cite{BC,HAD2} it is established that the  $d^*(2380)$ exists in the nuclear environment. The visible $d^*$ width is increased but consistent with the expected trivial effects of Fermi motion, the (known) increase of the $\Delta$ width in medium~\cite{ErWe} and also from additional open decay channels  such as $d^*(2380) N \rightarrow NNN$~\cite{MBH,EP,AP}. No significant change in the mass of the $d^*(2380)$ was obtained in medium from these experiments. We will therefore use the free space $d^{*}(2380)$ mass and width in our neutron star calculations, and modify these within realistic limits to explore the sensitivity for neutron star properties.
The limits we use are consistent with predicted modifications for dibaryons based on chiral symmetry restoration calculations \cite{DBT10}. We hope the  current work will provoke further studies in the area of modelling $d^*(2380)$ interactions in medium.
In any case the interactions between $d^*(2380)$ and nuclear matter may be expected to be weaker than for nucleons.\footnote{Due to the isoscalar nature of the $d^*(2380)$, isovector meson exchanges between the $d^*(2380)$ and nucleons are prohibited. Therefore $\pi$ and $\rho$ meson exchanges, the strongest contributor to NN-forces, are excluded and higher mass meson exchanges (e.g. the $\eta$, $\eta$' mesons) are known to have weak coupling to nucleons. The most likely mechanism of $d^*(2380)$ interaction with matter would be expected to be from $\sigma$-meson exchange (scalar-isoscalar two-pion exchange). From consideration of the relative coupling of the $\sigma$ with nucleons and $d^*(2380)$ it may be expected that even the $\sigma-d^*$ contribution would be strongly suppressed.} 

In infinite nuclear matter, as appropriate in neutron stars, the $d^*(2380)$ exists in a very different quantum mechanical environment than atomic nuclei. This can have important consequences such as the possible formation of a Bose-Einstein condensate of $d^*(2380)$ in the interior of neutron stars, analogous to the ones already identified for other exotic phases of bosons such as pions and kaons~\cite{Rho1}. Where the sum of neutron and proton chemical potentials are equal to that of the $d^*(2380)$ it  becomes energetically favorable for the system to store baryons in form of $d^*(2380)$, making the $d^*(2380)$ population stable in nuclear matter and producing an abrupt cut in neutron Fermi-momentum at $p^{Fermi}_{n}\sim 450 MeV$. For the typical separations of $d^*(2380)$ in neutron stars it can be assumed as a first approximation to be a point-like particle. We have used this assumption in our calculations.


\section{Effect of the $d^*(2380)$ dibaryon on neutron stars.}
The effect of dibaryons on the EoS for nuclear matter has been considered in various theoretical investigations, see {\it e.g.,} Refs.~\cite{NM1,NM2,NM3,NM4}. However such studies are generic and do not study a candidate identified experimentally, as in this present study. To study the impact of $d^*(2380)$ on the  composition, EoS and structure of neutron stars, we start from a
commonly used approach based on a relativistic Lagrangian describing the interaction of nucleons by means of the exchange of $\sigma-$, $\omega-,$ and $\rho-$mesons plus a free Fermi gas of leptons ($e^-$ and $\mu^-$). In particular, we use the GM1 parametrization of the Glendenning-Moszkowski model ~\cite{GM3}. This model predicts a maximum mass of $\sim 2.4 M_\odot$ for a pure nucleonic star, a value compatible with the current measurements of heavy neutron stars~\cite{Star1,Star2}. The EoS resulting from this choice of potential is also compatible with the latest stringent constraints on the radius of the 1.4$M_{\odot}$ star, extracted from the recent merger observation with gravitational waves \cite{Annala}. These valuable new data did not support the validity of NN potentials with a stiffer EoS. It would be premature to carry out a broader study including all existing NN potentials, as the $d^*(2380)$ would need to also be included consistently for the NN potential itself.\footnote{The $d^*(2380)$ has recently been shown to have a significant impact on the basic NN interaction in free space, even influencing the total cross section ~\cite{MBE1,MBE2}, effects which are not included in any current NN potential. Future calculations would therefore need to modify the existing NN-potentials incorporating $d^*$-mediated NN interactions as an explicit degree of freedom. We should also remark that the $d^*(2380)$ may also influence 3N and 4N forces.}

Although the $d^*(2380)$ interaction in matter is thought to be weaker than for nucleons and no significant modification is observed in nuclear matter, we presently lack detailed constraint of its interaction with other particles. Quark models tend to predict some attraction between the $d^*$ and the nuclear matter, while in molecular picture the $d^*$ interaction might become repulsive at higher densities ~\cite{AGPr}.
Therefore, in this work we include it just as a free gas of pointlike bosons. Clearly, the inclusion of interactions between the $d^*(2380)$ and nucleons or mesons should be a focus for future work when experimental information becomes available. To explore sensitivities to in-medium modifications of the $d^*(2380)$ in our model, we have varied its mass by $\pm 100$~MeV, in an attempt to asses the effect of an attractive or repulsive interaction.

\begin{figure}[!h]
\begin{center}
\includegraphics[width=0.5\textwidth,angle=0]{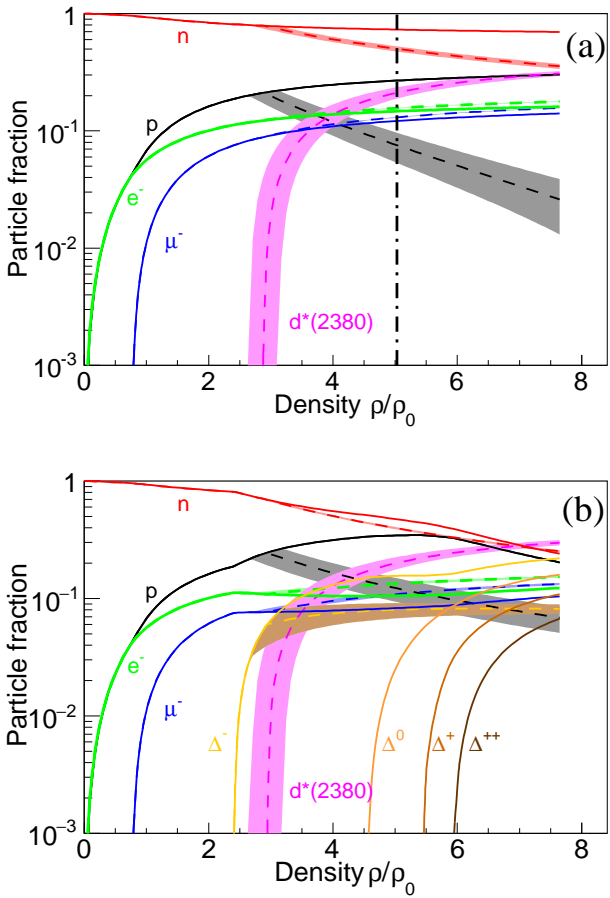}
\end{center}
\caption{(Color online) Particle fractions as a function of baryon density in units of saturation density $\rho_0$. The predictions assuming $m_{d^*}=[2280,2480]$ MeV and $m_{d^*}\equiv 2380$ MeV are shown by the shaded bands and dashed lines respectively, without explicit $\Delta$ degrees of freedom (model (a), top) and with (model (b), bottom). The vertical dash-dot line indicate the density of a heaviest possible neutron star for a nominal $d^*(2380)$ mass (see Table 1 for details).}
\label{composition}
\end{figure}

In Fig.\ref{composition} model(a), we show the predicted change in the chemical composition of a neutron star when we include the $d^{*}(2380)$ resonance as a free particle (dashed lines). The variation of the predictions with an in-medium $d^{*}$ mass modification in the range $m_{d^*}$ $[2280,2480]$ is shown by the shaded band.  For comparison the composition when only nucleons and leptons are considered is also presented (solid lines). The $d^*$ is predicted to appear at densities between 2.7 and 3.2 times normal nuclear matter saturation density ($\rho_0=0.16$ fm$^{-3}$). Note that since the $d^*(2380)$ has baryon number 2 its appearance induces an important and significant reduction in the neutron and proton fractions.  In addition, since it is positively charged,  the $e^-$ and $\mu^-$ fractions increase in order to maintain charge neutrality.

We also explored possible competition of the $d^*(2380)$ with other postulated constituents of nuclear matter. As the $d^*(2380)$ is strongly coupled to $\Delta\Delta$, in Fig.~\ref{composition}~(b) we show results where we additionally included the $\Delta$ quartet as an explicit degree of freedom in our model. The mass of the $d^*(2380)$ is around 80 MeV below the pole mass of two $\Delta$'s and therefore would tend to be preferentially created in nuclear matter. The one exception is the $\Delta^{-}$ which due to its favourable negative charge, appears at lower densities than the other members of the multiplet. However, in general the main conclusions for the important role of the $d^*(2380)$ remain unchanged with inclusion of the $\Delta$'s and in fact the predicted $d^*$ fraction at higher densities is increased.

\begin{figure}[!t]
\begin{center}
\includegraphics[width=0.5\textwidth,angle=0]{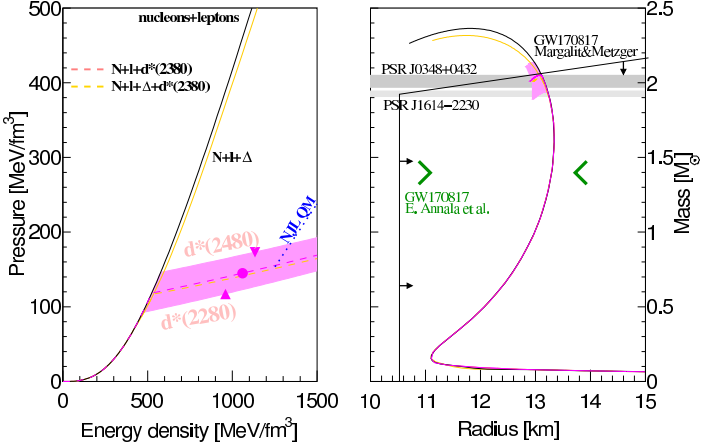}
\end{center}
\caption{(Color online) Neutron star EoS (left) and mass-radius relation (right) with and without the $d^*$ degree 
of freedom. The predictions assuming $m_{d^*}=[2280,2480]$ MeV and $m_{d^*}\equiv 2380$ MeV are shown by the shaded bands and dashed pink lines respectively. The effect of $\Delta$ degree of freedom shown as gold line with $d^*(2380)$(dashed) and without (solid). The observational masses of the pulsars PSR J1614-2230 ($1.928 \pm 0.017 M_\odot$) \cite{Star1} and PSR J0340+0432 ($2.01\pm 0.04 M_\odot$) \cite{Star2} as well as neutron star merger GW170817 limits from \cite{Margalit} and \cite{Annala} are also shown. The pink markers on a left panel represent maximum achievable pressure/energy density for heavy neutron star with $d^*$ degrees of freedom for the $m_{d^*}= 2380$ MeV(circle), 2280 MeV and 2480 MeV(triangles). Pure NJL quark matter EoS from Ref.\cite{NJL} is shown by dashed blue line.}
\label{mass}
\end{figure}

The neutron star EoS and the corresponding mass-radius relationship (Fig.~\ref{mass}), obtained by solving the Tolmann--Oppenheimer--Volkoff equations \cite{TOV1, TOV2}  is observed due to the bosonic character of the $d^*(2380)$ and to the reduction of the neutron and proton fractions. As a consequence the mass of the star is reduced from about $2.4 M_\odot$ to values in the range $1.9-2.1 M_\odot$.

The EoS for nucleonic degrees of freedom alone illustrates the expected relationship between pressure and energy density - increasing the pressure results in a rapidly and continuously increasing energy density. However, the $d^*(2380)$ offers new opportunities for the matter to respond to pressure increases. With the inclusion of the $d^*$ a significantly different behaviour is evident. The matter is predicted to undergo a phase transition which creates a much stronger correlation between pressure and energy density and limits the maximum achievable pressure. The upper limits for the central energy desnities of stable neutron stars in our model are shown by the markers on the figure, illustrating that such effects may contribute even within the mass range of stable neutron stars.

The mass-radius relationship predicted by our model is shown in the right panel. The nucleonic and nucleonic+$\Delta$ predictions give similar loci. However the predictions with the $d^*(2380)$ show an abrupt halt to the mass-radius locus at around $2 M_\odot$. The recent LIGO observation of a neutron star merger \cite{Abbott} allows a strict upper limit to be placed on the maximal neutron star mass of $M_{NS}<2.17M_{\odot}$ \cite{Margalit}. It is interesting to note that the cutoff produced by the $d^*(2380)$ phase transition in our model is in agreement with this stringent constraint. Our dimensionless tidal deformability parameter $\Lambda_{M_{NS}=1.4M_{\odot}}=740$ is also consistent with LIGO~\cite{Annala}. The maximum observed star mass from x-ray observations \cite{Star1,Star2} is also consistent with the mass cutoff predicted here. 
The predicted maximum stable neutron star mass, radius, density and $d^*$ fraction are presented in Table 1. The $d^*$ fraction is predicted to be around 20\% at the center of the star with maximum mass and, compared to the nucleonic case, the maximum star radius is increased and the central density reduced.

\begin{table}
\caption{Maximum mass neutron star properties}
\label{tab:1}       
\resizebox{0.65\textwidth}{!}{\begin{minipage}{\textwidth}
\begin{tabular}{lllll}
\hline\noalign{\smallskip}
Model & $M^{max}_{\odot}$ & Radius [km] & Central density $[\rho_{0}]$ & $d^*$ fraction at the center [\%] \\
\noalign{\smallskip}\hline\noalign{\smallskip}

Pure Nucleonic & 2.36 & 11.8 & 5.80 & 0 \\
Nucl.+ $d^*(2480)$ & 2.14 & 12.94 & 5.23 & 20 \\
Nucl.+ $d^*(2380)$ & 2.05 & 13.06 & 5.03 & 21 \\
Nucl.+ $d^*(2280)$ & 1.94 & 13.17 & 4.72 & 22 \\

\noalign{\smallskip}\hline
\end{tabular}
\end{minipage} }
\end{table}

 These first results indicate the $d^{*}(2380)$ has the possibility to play a signifivant role in neutron stars and dense cold nuclear matter. The interplay of the $d^{*}(2380)$ with other postulated degrees of freedom in high density nuclear matter should be investigated in future work.
The role of individual $\Delta$'s in dictating neutron star mass limits and properties appears to be diminished when the the $d^*(2380)$ is included in our model. Future work also investigating the interplay with hyperons would be a valuable advance. We note that previous work ~\cite{Drago1, Drago2} has indicated that the onset of hyperons is shifted to much higher densities ($\sim 5 \rho_0$) when the $\Delta$ is included. and similar effects may be expected with the $d^*(2380)$ contribution. At even higher densities transitions to deconfined quark matter are predicted \cite{NJL}. It would be interesting to investigate the influence of a bosonic $d^*(2380)$ in a transition to quark matter. The locattion of the predicted \cite{NJL} quark transition is also shown in Fig.~\ref{mass} left.

\section{The $d^*(2380)$ in star cooling and mergers.}
By comparing the predicted particle fractions with and without the $d^{*}(2380)$ (Fig.~\ref{composition}) it can be inferred that the majority of $d^{*}(2380)$ are produced from pn-pairs. The majority, but not all. Since the presence of the $d^{*}(2380)$ also increases the lepton (electron and muon) fraction, one can conclude that some of $d^{*}(2380)$'s would be produced in a weak process $nn\to d^*e\bar\nu_e$ or $nn\to d^*\mu\bar\nu_{\mu}$. This process would be a new, unanticipated mechanism for neutron star cooling. The anti-neutrinos produced in such weak $d^{*}(2380)$ formation would carry away energy right from the core of a neutron star. Further, the convection of $d^{*}(2380)$'s within the neutron star could also provide a new mechanism for star cooling. In regions of the star having densities below  $\sim2.8\rho_0$,  the $d^{*}(2380)$ would become unstable to $d^*\to nn e^+\nu_e$ decay. These neutrinos could also carry away energy in a dibaryon-Urca way. The main sites of these $d^{*}(2380)$ catalysed cooling processes would be rather non-uniform within the star. The antineutrinos would be produced from a spherical volume near the star centre, where the density is higher than $3\rho_0$, while neutrinos would be produced from a thin spherical surface at $\sim2.8\rho_0$.

In this first theoretical study we did not consider the effect of the $d^*(2380)$ spin on nuclear matter. The particle has the highest known spin for any hadron in its ground state, $J^{\pi}$=3$^{+}$.  One should therefore consider that the formation of the $d^{*}(2380)$ dibaryon does not only condense two baryons into the space of one, but also takes two units of angular momentum from the neutron star. Due to large factor $M_d^{*}/m_q\sim 7.6$ (with $m_q$ denoting constituent quark mass), the magnetic moment of a $d^*$ is expected to be very large, $\mu_d^{*} = 7.6 \mu_N$ \cite{Dong2017}. The modelling of this effect on the rotational dynamics  and magnetic field maps of neutron stars would be an important next step.

Considering that the $d^*(2380)$ is predicted to have a significant influence on the EoS at high densities (Fig.~\ref{mass}) further investigation of its role in neutron star merger and neutron star black hole mergers processes would be valuable. For example, the presence of significant $d^*(2380)$ fraction could influence the ejecta in star mergers. Although the maximum densities found in stable stars are expected (in our model) to be limited to around $5 \rho_0$, the densities in mergers would be much higher~\cite{NSML}, with the possibility of enhanced roles for the $d^*(2380)$ in the collision dynamics.  

Further, the decay of the $d^*(2380)$ in delocalised ejecta matter may offer the possibility to give additional high energy gamma production mechanisms. The kinetic energies of the nucleons inferred from the gamma ray burst spectra indicate the energies to produce the $d^*(2380)$ in NN collissions is easily reached  \cite{Fermi95} and the $\pi^0\pi^0 \to 4\gamma$ and $d^* \to d\gamma$ would both produced 100's MeV gammas in the d* rest frame.The inclusion of possible $d^*(2380)$ contributions into gamma ray burst simulation codes would be an interesting option. Further study to assess the possibility of detecting astronomical signatures of $d^*(2380)$ during mergers should also be investigated.  The detection of close to monoenergetic photons from the $d^*(2380)\rightarrow d\gamma$ decay would be a clear signal.

For nuclear matter out of equilibrium, as would be expected in the shock waves during a merger process, exotic routes may be followed to reach stable particle fractions. One such route is $d^*(2380)$ induced $n\to p$ conversion: $nn\to d^* e^- \bar\nu_e$. The characteristic timescale for this process to occur can be roughly estimated with the Fermi golden rule, used to evaluate neutron, muon, taon, etc lifetimes. According to Sargent's Law  the rate of this process is proportional to excess energy $Q^5$. In our case Q can be evaluated as a energy density difference between the $d^*(2380)$ and the pure nucleon EoS at the same pressure multiplied by the $d^*(2380)$ volume. The $d^*(2380)$ appears at a density $\rho=2.8\rho_0$ or at a pressure of 115 $MeV/fm^3$. At this density/pressure the conversion time from two neutrons to a $d^*$ would be essentially infinite compare to the time scales of a merger. However a slight increase in density to $3.0\rho_0$ or pressure to 120 $MeV/fm^3$ would decrease the conversion time below the free muon lifetime. And at $5-6\rho_0$ this process would proceed faster than the free tau decay. This is fast enough to have an effect on $d^*(2380)$ production in neutron star merger shock waves, in addition to the more trivial $pn \to d^*$ fusion.
 

\section{Summary and Conclusion}
We have evaluated the effect of the $d^{*}(2380)$ dibaryon on the nuclear equation of state and the mass-radius relation for neutron stars. The calculations used a simple bosonic gas approach for the $d^*(2380)$ supplementing a nucleonic equation of state calculated in a relativistic mean field approach. In our calculations the appearance of the $d^{*}(2380)$ dibaryons in nuclear matter limits the maximum possible neutron star mass to be around 2 solar masses, consistent with current observation limits and having an abrupt cut-off compatible with recent gravitational wave observations. The results indicate the $d^{*}(2380)$ could potentially be a new degree of freedom in neutron stars. Fractions of $d^*(2380)$ of around 20\% are predicted in the center of heavy stars, resulting in an increased maximum star radius and a reduced central density. New neutrino and antineutrino cooling mechanisms are possible with  $d^{*}(2380)$ formation, which have previously not been included in neutron star modelling.

These first results indicate that the $d^*(2380)$ is worthy of further investigation theoretically and experimentally to better constrain its role in neutron stars and star mergers.


\section{Acknowledgement}
We acknowledge valuable discussions with A. Gal. This work has been supported by STFC (ST/L00478X/1, ST/L005824/1) and by ``NewCompstar'', COST Action MP1304.


\end{document}